\documentclass[%
 reprint,
 amsmath,amssymb,
 aps,
prd,
showpacs
]{revtex4-2}

\usepackage{braket}
\usepackage{graphicx}% Include figure files
\usepackage{dcolumn}% Align table columns on decimal point
\usepackage{bm}% bold math
\begin{document}

%\preprint{APS/123-QED}

\title{Ideal gases and degenerate Fermi gases in external torsion fields}% Force line breaks with \\

\author{Chih-Hung Wang}

 \email{Corresponding author: robbin1101@gmail.com}

\author{Yu-Huei Wu}%
 \email{yuhueiwu@gmail.com}
\affiliation{%
Spacetime Academy, Taichung 404, Taiwan, ROC
}%


%\affiliation{
%\affiliation{
%\affiliation{%
%             %  but any date may be explicitly specified

\begin{abstract}

We investigate the effects of external torsion fields on ideal gases and Fermi gases, and derive a macroscopic quantity, which we call torsion susceptibility. We first consider the Dirac fermions in the Riemann-Cartan spacetime minimally coupled to the background torsion and electromagnetic fields. After applying the Foldy-Wouthuysen transformation, Hamiltonian of a spin-1/2 particle in weak field limit is obtained. The coupling of spin and spatial components of axial torsion vector has a Zeeman-like effect, which removes the degeneracy of energy levels and splits the energy levels with respect to the spin.  We calculate the macroscopic effects of the spin-torsion coupling on ideal gases, which satisfying the Boltzmann distribution, and Fermi gases, which satisfying the Fermi-Dirac distribution. The torsion susceptibility of ideal gases is inversely proportional to the temperature and is constant in Fermi gases. 
%\begin{description}
%\end{description}
\end{abstract}


\maketitle

%\tableofcontents

\section{\label{sec1}
Introduction
%First-level heading:\protect\\ The line
%break was forced \lowercase{via} \textbackslash\textbackslash
}

General relativity (GR) is a geometric theory of gravity based on the Riemannian geometry. The spacetime geometry is completely characterized by metric tensor and the sources of gravitational field are described by energy-momentum tensor.   The Einstein-Cartan (EC) theory is a natural extension of GR to spacetime geometry with torsion, i.e. Riemann-Cartan (RC) spacetime, where the metric tensor and metric-compatible connection are considered as  independent variables. In the EC theory, the spin tensor becomes the source of torsion fields. If we substitute the spin tensor into the torsion parts of the metric equations, the spin tensor parts in the metric equations can be interpreted as additional contribution to energy-momentum tensor. 

Besides the EC theory, torsion fields also plays an important role in the Poincar\'e gauge theory of gravity (PGT)\cite{Hehl-76}, where torsion is identified as the translational gauge field. In the PGT, the local gauge group is the Poincar\'e group, which is a semidirect product of the translation group times the Lorentz group, so gravitational fields are orthonormal co-frames $\{e^\alpha\}$ (translational gauge potentials) and metric-compatible connection with torsion $\{\omega^\alpha{_\beta}\}$ (rotational gauge potentials). The Noether theorem yields two gauge identities, which can be considered as equations of motion of matter fields \cite{Tucker-04, Wang-06}.

The equations of motion of classical spinning particles or fluids in the RC spacetime were studied in  \cite{Kopczynski-86, Obukhov-87}.  Yasskin and Stoeger \cite{YS-80} studied pole-dipole equations of motion of extended bodies in the RC spacetime, and found that torsion only couples to intrinsic spin (see also \cite{PO-07}).  The natural couplings of spin and torsion fields become a significant role on the detection of the spacetime torsion. Spin-polarized bodies have been used to constraint torsion fields in the laboratory experiments \cite{KRT-08, LYS-14} (also see a review article \cite{Ni-09}) . However, it is unclear to us how does microscopic spin-torsion couplings influence the macroscopic quantities of materials.  

The classical theory of paramagnetism and diamagnetism was first proposed by Langevin. However, the full descriptions of paramagnetism and diamagnetism require quantum mechanics. The microscopic theory of magnetism is based on quantum mechanics of electronic orbital angular momentum and intrinsic spin. In ideal gases, the intrinsic spin operator $\mathbf{\hat{s}}$ and orbital angular momentum operator  $\hat{\boldsymbol{{\mathit{l}}}}$ coupling to a uniformly magnetic field at microscopic scale changes the macroscopic magnetic moment of the ideal gases. The Boltzmann distribution allows us to calculate magnetic susceptibility $\chi$, which yields a positive susceptibility (paramagnetism) and inversely proportional to temperature $T$ (Curie's law) in weak magnetic field approximation (see Refs \cite{Landau-69, Coey, Kittle}). For a degenerate electron gas, which obeys Fermi-Dirac distribution, in weak magnetic fields, the intrinsic spin gives paramagnetic contribution (Pauli paramagnetism) and quantization of the orbital motion yields diamagnetic contribution (Landau diamagnetism).

Any interaction at the microscopic scales may change macroscopic quantities.  Statistical physics offers a path from microscopic interactions to macroscopic phenomena. It motivates us to calculate macroscopic quantities of materials from the microscopic interactions of gravitational fields and a Dirac particle. In the present work, we focus on the spin-torsion coupling and neglect any gravitational effect from metric tensor, so the background gravitational fields are considered as uniformly axial torsion vector with Minkowski metric tensor $ \eta$. 
%  It is possible to generalize our present work to consider other coupling terms of gravitational fields and Dirac particles. 

Dirac particles in the RC spacetime and background electromagnetic fields with  minimal and non-minimal couplings has been studied, either in the theoretical analysis or experimental constraints \cite{OST-15, CLR-21}. In Sec. \ref{sec2}, we give a brief review and present our notations and conventions.  In Sec. \ref{sec3}, we perform Foldy-Wouthuysen transformation \cite{BD-64, Silenko-08}, which decouples the particle and anti-particle wave functions, and obtain Hamiltonian of a spin-1/2 particle in the non-relativistic limit and weak external field approximation. 
We review the magnetization of the free-electron model  in Sec. \ref{sec4} and discuss the paramagnetism and diamagnetism of  ideal gases and the degenerate Fermi gases \cite{Landau-69, Coey}. In Sec. \ref{sec4-1}, we calculate the \textit{macroscopic axial-torsion moment} $\mathbf{\overline{\Upsilon}}$ and obtain \textit{molecular torsion susceptibility} $\chi_{\Theta} $ of ideal gases. We consider the Fermi-Dirac distribution and derive \textit{torsion susceptibility per unit volume} $ \chi_{\Theta}^{v}$ of degenerate Fermi gases in Sec. \ref{sec4-2}.  %We  discuss our results  in Sec. \ref{sec5}.

%This paper is outlined as below. In Sec. \ref{2}, we  generalize the BD theory to RC space-times. The main difference between our work and Dereli \& Tucker's work \cite{DT-82} is that our BD action with torsion includes three irreducible pieces of quadratic torsion, which were not considered in \cite{DT-82}. {These quadratic torsion terms may be associated to kinematic energy of orthonormal co-frames $e^a$.} Adding these terms does not spoil the field equations as the second-order differential equations. Without introducing any symmetry of space-time, we obtain a general torsion solution completely determined by $\Phi$, where the Lagrangian of matter is assumed to be the potential $U(\sigma)$ of the inflaton field.  In Sec. \ref{2-1}, we substitute the torsion solution back to our original action, and obtain an effective action, which is equivalent to the original BD theory except that $\omega(\Phi)$ now is a function of $\Phi$ instead of the dimensionless parameter. In Sec. \ref{3}, we study field equations in the homogeneous and isotropic Universe, and  obtain analytic and numerical solutions of $a(t)$ and $\Phi(t)$ during the inflation.  Sec. \ref{4} gives a discussion and conclusion. In this paper, we use the units $c=\hbar=1$ and $8\pi G= M_{Pl}^{-2}$ \cite{unit}.
%

\section{\label{sec2}
Dirac fermions in Riemann-Cartan spacetime
%First-level heading:\protect\\ The line
%break was forced \lowercase{via} \textbackslash\textbackslash
}

A Dirac spinor $\psi$ is a column matrix with 4 components, which are correspond to the wave functions of a particle and an antiparticle. The Dirac equation, which minimally coupled to the background electromagnetic and gravitational fields, is given by 
\begin{equation}
i \hbar \gamma^{\mu} \left( \partial_{\mu}  - \frac{1 }{2} \omega_{\alpha \beta}(X_{\mu}) \sigma^{\alpha \beta} - \frac{i}{\hbar} \frac{e}{c}  A_{\mu}  \right) \psi -  m c \psi = 0, \label{eq1}
\end{equation}
where $\omega_{\alpha \beta}= -\omega_{\beta\alpha}$ are the metric-compatible connection 1-forms,  $X_{\mu}$ are orthonormal frames, and $A_{\mu}$ are the components of the electromagnetic gauge potential 1-form $A$ with respect to orthonormal co-frames  $e^{\mu}$ . The four Dirac matrices  $\gamma^{\mu}$ satisfy the anti-commutation relations $\gamma^{\mu} \gamma^{\nu}+\gamma^{\nu} \gamma^{\mu}=2\eta^{\mu\nu}$, where $\eta^{\mu\nu}=\text{diag}(1, -1, -1, -1)$. The Greek indices are raised and lowered in terms of the Minkowski metric $\eta_{\mu\nu}$ and $ \partial_{\mu} =X_{\mu}^{i}\,\frac{\partial}{\partial x_i}$. The matrices 
\begin{equation}
\sigma^{\alpha \beta}  \equiv \frac{1}{4} [\gamma^{\alpha}, \gamma^{\beta}]=\frac{1}{4}\left(\gamma^{\alpha}\gamma^{\beta}-\gamma^{\beta}\gamma^{\alpha}\right)
\end{equation} are the generators of the Lorentz group. It is not difficult to verify that Eq. (\ref{eq1}) is covariant under the gauge transformations of the local Poincar\'e group and 1-dimensional Abelian $U(1)$ group. 

The standard representation of the Dirac matrices is given by \cite{BD-64, Dirac}
\begin{equation}
 \gamma^{i}=
\begin{pmatrix}
    0 & \sigma^i    \\
    - \sigma^i & 0 
\end{pmatrix}, \hspace{0.5cm}
\gamma^0=\beta=
\begin{pmatrix}
    1 & 0   \\
    0 & -1 
\end{pmatrix},
\end{equation} where $i=1, 2, 3$ and $\sigma^i$ are the 2$\times$2 Pauli matrices. The spin of electrons only requires two components of wave functions, so the four components of the Dirac spinor actually contain the positive-energy solutions (the states of particles) and negative-energy solutions (the states of antiparticles) \cite{Dirac}. 

In Eq. (\ref{eq1}), both $ \gamma^{\mu}  \partial_{\mu}$ and $\gamma^{\mu}\omega_{\alpha \beta}(X_{\mu}) \sigma^{\alpha \beta}$ contain the interactions of the gravitational fields $\{ e^\alpha,\, \omega^\alpha_{\,\,\,\beta}\}$ and the Dirac spinor $\psi$.  However, the spin-torsion coupling only appears in  $\gamma^{\mu}\omega_{\alpha \beta}(X_{\mu}) \sigma^{\alpha \beta}$. In the RC spacetime, the torsion 2-forms $T^\alpha$ and the curvature 2-forms $R^\alpha_{\,\,\,\beta}$ are defined as
\begin{eqnarray}
T^\alpha& =& d e^\alpha + \omega^\alpha_{\,\,\, \beta} \wedge e^\beta =D e^\alpha,    \label{eq4} \\
R^\alpha_{\,\,\,\beta}&=&d  \omega^\alpha_{\,\,\, \beta} +  \omega^\alpha_{\,\,\, \gamma} \wedge  \omega^\gamma_{\,\,\, \beta}, %\hspace{0.8cm}
\end{eqnarray} 
where $D$ is the covariant exterior derivative.  From Eq. (\ref{eq4}), we can express $\omega_{\alpha \beta}$ in terms of $d e^\alpha$ parts, which corresponds to the Levi-Civita connection, and $T^\alpha$ parts, which corresponds to the contortion tensor \cite{BT-87}.
 
 The Hermitian Hamiltonian in the RC spacetime with the Schwinger gauge has been shown in Ref. \cite{OST-15}. It turns out that the spin-torsion coupling in the Hamiltonian only has axial torsion 1-form (the totally antisymmetric torsion), which is
\begin{equation}
\Theta=\star(T^\alpha \wedge e_\alpha),
\end{equation} where $\star$ denotes the Hodge map. 
By introducing the spin matrices $\mathbf{\Sigma}=\{\Sigma^1= i\gamma^2\gamma^3, \Sigma^2= i\gamma^3\gamma^1, \Sigma^3=i\gamma^1\gamma^2\}$, which are
\begin{eqnarray}
\Sigma^i=\begin{pmatrix}
  \sigma^i    &   0 \\
     0 & \sigma^i 
\end{pmatrix},
% \\
%\Sigma^2=i\gamma^3\gamma^1=\begin{pmatrix}
%  \sigma^2    &   0 \\
%     0 & \sigma^2 
%\end{pmatrix}\\
%\Sigma^3=i\gamma^1\gamma^2=\begin{pmatrix}
%  \sigma^3    &   0 \\
%     0 & \sigma^3 
%\end{pmatrix} 
\end{eqnarray} 
and the orthonormal co-frame in the Schwinger gauge
\begin{eqnarray}
e^0   =  V c\,dt, \hspace{0.5cm} e^i=W^i_{\,\,\,j} (d x^j -  K^j c\, dt),
\end{eqnarray} where the functions $V$, $K^j $, and $W^i_{\,\,\,j}$ may depend on the coordinates $x^\mu=\{ct, x^i\}$, the spin-torsion coupling in the Hamiltonian yields \cite{OST-15}
\begin{equation}
-\frac{\hbar\, c\,V}{4} \left( \mathbf{\Sigma} \cdot \mathbf{{\Theta}}+ \gamma^5 {\Theta}^0 \right),
\end{equation} where $\tilde{\Theta}= \tilde{\eta}(\Theta, -) =\mathbf{{\Theta}} +  {\Theta}^0 X_0$ is the metric dual of $\Theta$ and can be considered as the axial torsion vector. The $4\times 4$ matrix  $\gamma^5=i \gamma^0 \gamma^1  \gamma^2  \gamma^3$.

\section{\label{sec3}
Spin-torsion coupling in the low-energy and weak field limits
}

The Dirac equation is a relativistic quantum equation for fermions, which exists the solutions of the negative energy. These solutions are corresponding to the quantum states of the anti-paritcles. In order to obtain sensible non-relativistic quantum equations for particles, one may need to decouple the Dirac equation into  two two-components equations. One two-components equation reduces to Schr\"odinger-like equations for a spin-1/2 fermion and the other describes the quantum states of a spin-1/2 anti-fermion. The Foldy-Wouthuysen transformation provides a systematic procedure to decouple the Dirac equation, so we can obtain the non-relativistic  Hamiltonian operator of the spin-1/2 particle in the external fields.

\subsection{\label{sec3-1}
The Foldy-Wouthuysen transformation for Dirac fermion
}

The operators in the  Hamiltonian of the Dirac fermions are classified into the "odd" operators which couple the large to small components, and the "even" operators  which do not couple those components \cite{BD-64}. The Foldy-Wouthuysen (FW) transformation is the unitary transformation trying to remove the odd operators in the Hamiltonian. Here, we omit the detail derivations of the FW transformation for Dirac fermion in the external electromagnetic and gravitational fields, which can be found in Ref. \cite{Silenko-08, OST-15}, and only discuss the physical meanings of the coupling terms in the non-relativistic Hamiltonian of the spin-1/2 particle.  

We start from the following Hamiltonian  
\begin{equation}
\label{eq10}
H= c \,\boldsymbol{\alpha} \cdot \left(\boldsymbol{p} -\frac{e}{c}\boldsymbol{A}\right) + \beta m c^2 + e\Phi -\frac{\hbar\,c}{4} \left( \mathbf{\Sigma} \cdot \mathbf{{\Theta}}+ \gamma^5 {\Theta}^0 \right) ,
\end{equation} which minimally couples to the  background electromagnetic and axial torsion fields. $\boldsymbol{p}$ is the 3-dimensional momentum operator, and the matrices $\boldsymbol{\alpha}=\{\alpha^i\}$ are defined as $\alpha^i=\beta \gamma^i$. The electromagnetic vector potential $\tilde{A}=\tilde{\eta}(A, -)=\{\Phi, \boldsymbol{A}\}$. The metric fields do not appear in Eq. (\ref{eq10}) since we assume the background orthonormal co-frames is simply as $e^\alpha=\{d t,\, d x^i\}$, which corresponds to $V=1, \,W^i_{\,\,\,j}=\delta^i_{\,\,j},\, K^i=0$.

After applying FW transformation to Eq. (\ref{eq10}) and taking weak-field limit, we obtain 
\begin{eqnarray}
\label{eq11}
&&H_{FW}=\beta \left(m c^2 + \frac{\boldsymbol{\pi}^2}{2m} -  \mu_B\, \boldsymbol{\Sigma} \cdot \boldsymbol{B}  + \frac{\hbar}{8 m} \{\Theta^0, \boldsymbol{\pi} \cdot \boldsymbol{\Sigma}\} \right)   \nonumber \\
 &&+ \,e\Phi -\,\frac{\hbar c}{4} \mathbf{\Theta} \cdot \boldsymbol{\Sigma} - \frac{e\hbar}{m^2 c^2} \boldsymbol{\Sigma} \cdot \left( \boldsymbol{\pi}\times \boldsymbol{E} - \boldsymbol{E}\times \boldsymbol{\pi} \right) ,
\end{eqnarray} where $\boldsymbol{\pi}=\boldsymbol{p} -\frac{e}{c}\boldsymbol{A}$, and curly brackets $\{\,\, ,\,\, \}$ denote anti-commutations. $\mu_B=\frac{e\hbar}{2mc}$ is the Bohr magneton. $\boldsymbol{E}=\{E^i\}$ and $\boldsymbol{B}=\{B^i\}$ are electric and magnetic fields. We neglect the orders higher than $O(1/m^2)$ terms  in the $H_{FW}$. 

The $\boldsymbol{\pi}^2$ term contains the coupling terms of orbital angular momentum and the magnetic field, which will be discussed in Sec. \ref{sec3-2}.  The $e\Phi$ term is the electrostatic energy and the $\boldsymbol{\Sigma} \cdot \boldsymbol{B}$ term is magnetic dipole energy of the intrinsic spin.  Actually, $\boldsymbol{\Sigma} \cdot \boldsymbol{B}$ term yields the paramagnetism in the magnetization of the ideal gases and degenerate Fermi gases. The last term in $H_{FW}$ contains the spin-orbital coupling $\mathbf{\hat{s}} \cdot \boldsymbol{\hat{{\mathit{l}}}}$, which is related to the fine structure of atomic levels. 

We first notice that the coupling of the spatial axial torsion $\mathbf{\Theta}$ and $\boldsymbol{\Sigma}$  has the same form as $\boldsymbol{\Sigma} \cdot \boldsymbol{B}$, so we expect that the direction of the intrinsic spin $\hat{\mathbf{s}}$ of the spin-1/2 particles should line up along the direction of $\mathbf{\Theta}$. We call this effect \textit{paratorsionism}. In Sec. \ref{sec4}, we will  investigate the macroscopic effects of the interaction $\mathbf{\Theta} \cdot \boldsymbol{\Sigma}$ and the effects of the $\Theta^0$ coupling term may be discussed in the future work. 

%\subsubsection{Wide text (A level-3 head)}
%The \texttt{widetext} environment will make the text the width of the
%full page, as on page~\pageref{eq:wideeq}. (Note the use the
%\verb+\pageref{#1}+ command to refer to the page number.) 
%\paragraph{Note (Fourth-level head is run in)}
%The width-changing commands only take effect in two-column formatting. 
%There is no effect if text is in a single column.

\subsection{\label{sec3-2}
A spin-1/2 particle in the uniform magnetic and torsion fields.  
%Citations and References
}
Let us consider a simple situation where the background fields are uniform magnetic and spatial axial torsion fields, i.e. both $\boldsymbol{B}$ and $\mathbf{\Theta}$ are constants. This consideration may not be unreasonable since external constant magnetic field has been used to study magnetization of materials  in the laboratory experiments.  The spatial vector potential $\boldsymbol{A}$ in the uniform magnetic field satisfies the form 
\begin{equation}
\label{ }
\boldsymbol{A}=\frac{1}{2} \, \boldsymbol{B} \times \boldsymbol{r},
\end{equation} 
so a straightforward calculation yields the reduced Hamiltonian $\mathcal{H}$ of Eq. (\ref{eq11})
\begin{eqnarray}
\label{eq13}
\hat{\mathcal{H}}&=& \left(mc^2 + \frac{\boldsymbol{\hat{p}}\cdot \boldsymbol{\hat{p}} }{2m}\right) - \mu_B (\boldsymbol{\hat{\mathit{l}}}+ 2\,\mathbf{\hat{s}})\cdot \boldsymbol{B}-\frac{\hbar c}{2} \mathbf{\hat{s}} \cdot \boldsymbol{\Theta} \nonumber\\
&&+\,\frac{e^2}{8mc^2}\,\,( \boldsymbol{B} \times \boldsymbol{r}) \cdot ( \boldsymbol{B} \times \boldsymbol{r}),
\end{eqnarray} where we put $\hat{ }$ on the physical quantities to denote that these quantities are operators.  The orbital angular momentum operator $\hat{\boldsymbol{{\mathit{l}}}}$ and the intrinsic spin operator $\mathbf{\hat{s}}$ are defined as
\begin{equation}
\label{ }
\hbar\,\boldsymbol{\hat{\mathit{l}}} =  \boldsymbol{r} \times {\boldsymbol{\hat{p}}}, \hspace{0.5cm} 
\mathbf{\hat{s}}=\frac{1}{2}\, \boldsymbol{\hat{\sigma}},
\end{equation} 
where $\boldsymbol{\hat{\sigma}}=\{\sigma^i\}$. We should point out that Eq. (\ref{eq13}) does not include the negative-energy states. 

It is convenient to introduce the magnetic moment $\boldsymbol{\hat{m}}=-\mu_B\, g \boldsymbol{\hat{J}} $, where $\boldsymbol{\hat{J}} =\boldsymbol{\hat{\mathit{l}}} +\mathbf{\hat{s}}$ is the total angular momentum and 
\begin{equation}
\label{ }
g=1\,+\frac{J(J+1)-l(l+1)+s(s+1)}{2J(J+1)} 
\end{equation} is called the Land\'e factor or the gyromagnetic factor \cite{Landau-65}. If there is no orbital angular momentum in the unperturbed states, i.e. $l=0$, we obtain $g=2$, which is predicted by the Dirac equation. If we place an atom in a uniform magnetic field,  the second term of Eq. (\ref{eq13}), which is magnetic dipole energy $\boldsymbol{\hat{m}}\cdot \boldsymbol{B}$, splits the atomic levels and removes the degeneracy with respect to the directions of the total angular momentum. This phenomena is called the \textit{Zeeman effect}. We can also define the axial-torsion moment $\boldsymbol{\hat{\Upsilon}}= -\,\frac{\hbar c}{2}\, \mathbf{\hat{s}}$, so the axial-torsion dipole energy, $\boldsymbol{\hat{\Upsilon}}\cdot \boldsymbol{\Theta} $ in Eq. (\ref{eq13}) also yields a \textit{Zeeman-like effect}.

%
%A citation in text uses the command \verb+\cite{#1}+ or
%\verb+\onlinecite{#1}+ and refers to an entry in the bibliography. 
%An entry in the bibliography is a reference to another document.

\section{\label{sec4}
Macroscopic effects of spin-torsion coupling on ideal and Fermi gases
}
The magnetization of the materials has been well studied (see Refs. \cite{Coey, Kittle}). The phenomena of the paramagnetism and diamagnetism originally come from the microscopic interactions of the second and fourth terms of Eq. (\ref{eq13}). The  distribution of the microscopic states of a macroscopic subsystem satisfies the grand canonical distribution function \cite{Landau-69}
\begin{equation}
\label{eq16}
\mathrm{w}_{nN}=e^{(\Omega+\mu N-E_n)/k_BT},
\end{equation} which depends on energy eigenstates $E_n$ and the numbers $N$ of particles of the subsystem. $\Omega$ is the thermodynamical potential, $\mu$ is the chemical potential of the subsystem, and $k_B$ is the Boltzmann constant. The macroscopic quantities (expectation values) of the subsystem can be calculated from Eq. (\ref{eq16}). If an external field changes the Hamiltonian of an atom or a molecule through the couplings, the distribution function $\mathrm{w}_{nN}$ will also changes. It turns out that the macroscopic quantities may be modified.

 If we set the z-axis along the uniform magnetic field $\boldsymbol{B}$ and consider $\boldsymbol{\Theta}=0$, the expectation value per gas molecule $\overline{m}_z$ of the ideal gases, which satisfy the Boltzmann distribution $\mathrm{w}_k\propto e^{-\varepsilon_k/k_BT}$, yields \cite{Coey}
 \begin{equation}
\label{eq17}
\overline{m}_z=\frac{\sum_{-J}^{J} \mu_B\, g\, m_J \,e^{-x}}{\sum_{J}^{J} e^{-x}}=g\, \mu_B  J\, B_J(x),
\end{equation} where $m_J$ is the azimuthal quantum number, $B_J(x)$ is the Brillouin function, and  $x= g \mu_B J B/k_B T$.  In the small-$x$ limit, i.e. the weak $B$ field limit, the leading term of Eq. (\ref{eq17}) gives molecular magnetic susceptibility
 \begin{equation}
\label{eq18}
\chi=\frac{\overline{m}_z}{B}=\frac{\mu_B^2\,g^2\,J(J+1)}{3k_BT},
\end{equation}which satisfies the Curie's law. Since $\chi>0$, so it is paramagnetism. 

If we take Eq. (\ref{eq16}) and consider the Pauli's exclusive principle, it yields the Fermi-Dirac distribution
\begin{equation}
\label{eq19}
\overline{n}_k=1/( e^{(\varepsilon_k-\mu)/k_BT}+1),
\end{equation}
where $\overline{n}_k$ is the mean number of particles in the $k$th quantum state. When the temperature is near absolute zero, the distribution of the electron gas satisfies Eq. (\ref{eq19}), which we called degenerate electron gases.  The spin-magnetic coupling term ($\mathbf{\hat{s}}\cdot \boldsymbol{B}$) and orbital-magnetic coupling term ($\boldsymbol{\hat{\mathit{l}}} \cdot  \boldsymbol{B}$) in the degenerate electron gases yield the Pauli paramagnetism  and Landau diamagnetism, respectively \cite{Landau-69}. Since spatial axial torsion $\boldsymbol{\Theta}$ is only couple to intrinsic spin $\mathbf{\hat{s}}$ in Eq. (\ref{eq13}), we would expect spatial axial torsion only produce paratorsionism.

\subsection{\label{sec4-1}
Molecular torsion susceptibility of the idea gases
%Citations and References
}
We start from an ideal gas, which has total number of particles  $N$, in the background uniform spatial axial torsion field  $\boldsymbol{\Theta}$ and $\boldsymbol{B}=0$, and set the z-axis along the direction of $\boldsymbol{\Theta}$. In the weak $\Theta_z$ limit,  the spin-torsion coupling in Eq. (\ref{eq13}) spilts the unperturbed energy $\varepsilon^{(0)}_k$ and yields 
\begin{equation}
\label{eq20}
\Delta \varepsilon_k\equiv\varepsilon_k-\varepsilon^{(0)}_k=- \mathcal{A}_k \Theta_z -\mathcal{B}_k \Theta_z^2,
\end{equation}  where
\begin{equation}
\label{ }
\mathcal{A}_k=\braket{k|\hat{\Upsilon}_z|k}, \hspace{0.3cm}  \mathcal{B}_k =  \sum_{k'} \frac{|\braket{k|\hat{\Upsilon}_z|k'}|^2}{\varepsilon^{(0)}_{k'}-\varepsilon^{(0)}_k}\, (k'\neq k).
\end{equation} Here,  $\ket{k}$ corresponds to unperturbed energy eigenstate. 

If we substitute $\varepsilon_k$ into the partition function $Z=\sum_k e^{-\varepsilon_k/k_B T}$ and expand the exponential function in terms of $\Theta_z$, it yields 
\begin{equation}
\label{eq21}
Z=\left( 1+\frac{\overline{\mathcal{A}^2}\,\Theta_z^2}{2k_B^2T^2}+\frac{\overline{\mathcal{B}}\,\Theta_z^2}{k_BT}\right) \sum_k e^{-\varepsilon^{(0)}_k/k_B T},
\end{equation} where the bar denotes averaging over the Boltzmann distribution unperturbed by the field. The linear term $\overline{\mathcal{A}}$ in $Z$ vanishes because of the symmetry of the unperturbed  energy eigenstate. 

It is clear that a small change of $\delta\Theta_z$ will cause the Hamiltonian of the ideal gas change, so the external $\Theta_z$ field may play as an extra parameter in the free energy $F$ of the gas. We can write
\begin{equation}
\label{ }
(\delta F)_{T, V, N}= -\overline{\Upsilon}_z\, \delta \Theta_z,
\end{equation} where $T$ and $V$ denote temperature and volume of the gas, and the mean torsion moment is 
\begin{equation}
\label{ }
 \overline{\Upsilon}_z=-(\partial F/ \partial \,\Theta_z)_{T, V, N}.
\end{equation}
By substituting Eq. (\ref{eq21}) into the free energy 
\begin{equation}
\label{ }
F=-k_BTN\ln{Z} +k_BT \ln{N!} 
\end{equation}
and differentiating the free energy with respect to $\Theta_z$, we obtain the molecular torsion susceptibility 
 \begin{equation}
\label{ }
% \overline{\Upsilon}_z= N \Theta \,\chi_\Theta,
\chi_\Theta  =\frac{\overline{\Upsilon}_z}{N \Theta}= \frac{\overline{\mathcal{A}^2}}{k_BT} + \overline{\mathcal{B}}.
\end{equation}  %$\chi_\Theta$ yields
%\begin{equation}
%\label{ }
%\end{equation} 
Since $\overline{\mathcal{B}}$ term is belong to the second-order perturbation,  this effect will be analyzed in other works. 

We suppose that the temperature is small  in comparsion with the interval between the ground state and the nearest excited state, so only the ground state contributes to the mean value $\overline{\mathcal{A}^2}$. A straightforward calculation yields
\begin{equation}
\label{eq27}
\chi_\Theta=\frac{\hbar^2 c^2 s(s+1)}{12\,k_B T},
\end{equation} which is also inversely proportional to temperature as  the result in Eq. (\ref{eq18}).   Since $\chi_\Theta>0$, so it is paratorsionism. 

\subsection{\label{sec4-2}
Torsion susceptibility per unit volume of the degenerate elecron gases
%Citations and References
}
Let us now consider an electron gas in the low temperature $T$, which is the degenerate case. Actually, the condition for the electron gas to be the strongly degenerate is $k_B T\ll\varepsilon_F$,  where $\varepsilon_F$ is the Fermi energy of the degenerate electron gas.  Since number of particles in a given quantum state is not fixed, it is more useful to introduce the thermodynamic potential $\Omega=F - \mu \overline{N}$, where $\overline{N}$ is the expectation value of the subsystem and $\Omega$ is a function of $T, V$ and $\mu$. Similarly, $(\delta \Omega)_{T, V, \mu}=-\overline{\Upsilon}_z\, \delta \Theta_z $ yields the mean torsion moment 
\begin{equation}
\label{ }
\overline{\Upsilon}_z=-(\partial \,\Omega/ \partial \,\Theta_z)_{T, V, \mu}. 
\end{equation}  If we take the perturbation to first-order and put $s=\pm \frac{1}{2}$, Eq. (\ref{eq20}) yields $\varepsilon=\varepsilon^{(0)}\pm \frac{\hbar c}{4}\,\Theta_z$. The thermodynamical potential $\Omega$ of an electron gas in a uniform $\Theta_z$ field may be written 
\begin{equation}
\label{eq29}
\Omega(\mu)=\frac{1}{2}\,\Omega_{0}(\mu+\frac{\hbar c}{4} \Theta_z)+\frac{1}{2}\,\Omega_{0}(\mu-\frac{\hbar c}{4} \Theta_z),
\end{equation} where $\Omega_0$ is the potential in the absence of the torsion fields and the factor $ \frac{1}{2}$ denotes the half of the number of quantum states of the electron when the value of its spin is fixed. Expanding Eq. (\ref{eq29}) in powers of $\Theta_z$, we obtain
\begin{equation}
\label{ }
\Omega(\mu)\cong\Omega_0+\frac{\hbar^2 c^2}{32} \Theta_z^2\left(\frac{\partial^2 \Omega_0(\mu)}{\partial \mu^2}\right),
\end{equation} and use  $(\partial \,\Omega_0(\mu)/ \partial \mu)_{T, V}=-N$, it yields torsion susceptibility per unit volume  
\begin{equation}
\label{eq22}
\chi_\Theta^v=\frac{\overline{\Upsilon}_z}{V\Theta_z}=\frac{\hbar^2 c^2}{16V}\left(\frac{\partial \,N}{\partial \mu}\right)_{T, V}
\end{equation}

If we consider the temperature is near absolute zero ($T\to 0$), it gives $\varepsilon_F\simeq\mu$.  By substituting the total number of the electrons 
%\begin{equation}
%\label{ }
$N=\frac{V\sqrt{(2m\mu)^{3}}}{3\pi^2 \hbar^3}$
%\end{equation}
 into Eq. (\ref{eq22}), we have temperature-independent torsion susceptibility
\begin{equation}
\label{eq23}
\chi_\Theta^v=\frac{c^2}{32\hbar\pi^2}\left(\sqrt{8m^3\varepsilon_F}\right),
\end{equation} which is also paratorsionism.

\section{\label{sec5}
Discussions
}
We investigate the torsion effects on the ideal gases and the degenerate Fermi gases, and obtain the macroscopic torsion moment and torsion susceptibility, Eq. (\ref{eq27}) and (\ref{eq23}). Although the existence of the torsion fields have not been fully confirmed by the laboratory experiments or scientific observations, our work provides a new way to search the torsion effects.  These macroscopic phenomena, \textit{paratorsionism} or \textit{diatorsionism}, may either be observed in the laboratory or be used to constraint axial torsion on the Earth. 

A direct extension of our present work is to consider microscopic interactions of metric tensor and the Dirac particles and study the macroscopic effects of these interactions. There are other coupling terms in the Hamiltonian $H_{FW}$, which may give us interesting macroscopic phenomena. 

We are also curious the torsion effects in the neutron stars or the early Universe, so our next step is to generalize our work to the extreme relativistic Fermi gases. It is quite reasonable that torsion and matter-field couplings in the early Universe yields macroscopic quantities, which may imprint on astrophysical or cosmological observations. 

%\bibliography{apssamp}%

%\Sigma_{n=1}^{10}

%\blindtext \cite{article-minimal}

%\bibliographystyle{apsrev4-1} % Tell bibtex which bibliography style to use
%\bibliography{xampl}
\end{document}